\documentclass{elsart}

\usepackage{amssymb}
\usepackage{bm}
\usepackage{psfig}

\usepackage{natbib}

\newcommand{\tpartial}{\tilde\partial}
\newcommand{\tz}{\tilde{\bm z}}
\newcommand{\tnabla}{\tilde\nabla}
\newcommand{\tJ}{\tilde J}

\newif\iffigs
\figstrue 
\iffigs
\fi
\def\drawing #1 #2 #3 {
\begin{center}
\setlength{\unitlength}{1mm}
\begin{picture}(#1,#2)(0,0)
\put(0,0){\framebox(#1,#2){#3}}
\end{picture}
\end{center} }

\def\v{{\bm v}}

\def\x{{\bm x}}

\def\y{{\bm y}}
\def\z{{\bm z}}
\def\k{{\bm k}}
\def\e{{\bm e}}

\def\rf#1{(\ref{#1})}

\def\un{\hbox{{1\kern -0.25em\raise 0.4ex\hbox{{\scriptsize $|$}}}}}

\def\e{{\rm e}}
 
\begin{document}
\begin{frontmatter} 
\title{The analytic structure of 2D Euler flow at short times}
\author[kyoto,oca]{T.~Matsumoto\corauthref{cor}}
\corauth[cor]{corresponding author, email: takeshi@kyoryu.scphys.kyoto-u.ac.jp} 
\author[oca,ias]{, J.~Bec}
\author[oca,ias]{, U.~Frisch}
\address[kyoto]{\mbox{Dep. Physics, Kyoto University, Kitashirakawa Oiwakecho Sakyo-ku, Kyoto 606-8502, Japan}}
\address[oca]{\mbox{CNRS UMR 6529, Observatoire de la C\^ote d'Azur, BP 4229, 06304 Nice Cedex 4, France}}
\address[ias]{\mbox{Institute for Advanced Study, Einstein drive, Princeton, N.J. 08540, USA}}
\date{\today}
\vspace{4mm}
\centerline{\normalsize \textit{Fluid Dyn. Res.}~\textbf{36}, 221--237 (2005).}
\begin{abstract}
Using a very high precision spectral calculation applied to the 
incompressible and inviscid flow with
initial condition $\psi_0(x_1,\,x_2) = \cos x_1+\cos 2x_2$, we find that the
width $\delta(t)$ of its analyticity strip follows a $\ln(1/t)$ law at short
times over  eight decades. The asymptotic equation governing the
structure of spatial complex-space singularities at short times (Frisch,
Matsumoto and Bec 2003, J.~Stat.~Phys. 113, 761--781) is solved by a
high-precision expansion method. Strong numerical evidence is obtained that
singularities have infinite vorticity and lie on a complex manifold 
which is constructed explicitly as an envelope of analyticity disks.
\end{abstract}
\begin{keyword}
\PACS  47.11.+j 02.40.Xx 0230.Fn 02.70.Hm \\
Euler equation; asymptotics; complex singularities.
\end{keyword} 
\end{frontmatter} 
 
\section{Introduction}
\label{s:intro}

In early September 2001 one of the authors (UF) attended the Zakopane meeting on
Tubes, Sheets and Singularities \citep{zakopane} which was also attended by
Richard Pelz. There were many discussions about the issue of finite-time
blowup for 3D incompressible Euler flow.  Richard, who had studied a flow
introduced by \citet{kida85}, had evidence in favor of blowup but one could
not rule out that the highly special structure of this flow would lead to
quasi-singular intermediate asymptotics. The three authors of this paper then
decided to embark in a long-term project aimed at getting strong evidence for
or against blowup for a wide class of flows encompassing the Taylor--Green
flow \citep{brachetetal} and the Kida--Pelz vortex
\citep{kida85,p97,pg97,p02}, namely space-periodic flow with or without
symmetry having initially only a few Fourier harmonics. Such initial flow is
not only analytic but also entire: there is no singularity at finite distance
in the whole complex spatial domain.

As has been known since the mathematical work of \citet{bbz76}, any real
finite-time singularity is preceded by complex-space singularities approaching
the real domain and which can be detected and traced using Fourier methods
\citep{tracing, fmb03}.  This method is traditionally carried out by spectral
simulations which run out of steam when the distance $\delta(t)$ from the real
domain to the nearest complex-space singularity is about two meshes. We
pointed out in a recent paper \citep[][henceforth referred to as FMB]{fmb03},
which also reviews the issue of blowup, that it may be possible to
extend the method of tracing of complex singularities by performing a
holomorphic transformation mapping singularities away from the real domain
and, perhaps doing this recursively. This is the basic idea of the {\it
spectral adaptive\/} method which aims at combining the extreme accuracy of
spectral methods with the local mesh refinement permitted by adaptive methods.

In one dimension the complex-space singularities of PDE's are isolated points,
at least in the simplest cases, as for the Burgers equation.  In higher
dimension they are extended objects, such as complex manifolds.  Understanding
the nature and the geometry of such singularities is a prerequisite for
mapping them away. Many aspects can already be investigated in the
two-dimensional case for which we know not only that blowup is ruled out, but
we also know that the flow stays a lot more regular than predicted by rigorous
lower bounds (basically $\delta(t)$ seems to decrease exponentially
whereas the bound is a double exponential in time). In FMB we gave some
evidence that in 2D the complex singularities are on a smooth manifold, but
the nature of the singularities was not very clear and in particular the issue
of finiteness vs. blowup of the complex vorticity was moot.

In FMB we also pointed out that the issue of singularities can
be simplified if we limit ourselves to short times. Let us briefly
recall the setting. 
We start with the 2-D Euler equation written in stream function formulation
\begin{equation}
  \partial_t \nabla^2 \psi = J(\psi,\nabla^2 \psi),
  \label{streamform}
\end{equation}
where $J(f,g)\equiv \partial_1f\partial_2g
-\partial_1g\partial_2f$. As in FMB, we focus on the two-mode initial
condition
\begin{equation}
\psi_0(\x) = \cos (x_1) + \cos (2x_2),
\label{deuxmodes}
\end{equation}
one of the simplest initial condition having nontrivial Eulerian dynamics. 
The solution $\psi(\z,t)$, obtained by analytic continuation to
complex locations $\z=\x+i\y$, is expected to have singularities at large
imaginary values when $t$ is small. If one focuses on the quadrant 
$y_1 \to +\infty$ and $y_2 \to +\infty$, an
asymptotic argument given in FMB  suggests looking at solutions satisfying the 
\textit{similarity ansatz} 
\begin{eqnarray}
\psi(\z,t) &= &(1/t) F(\tz),
\label{simansatz}\\
\tz& =& (\tilde z_1,\, \tilde z_2) \equiv (z_1+i\ln t,\, z_2+(i/2) \ln t).
\label{deftz}
\end{eqnarray}
Substitution in \rf{streamform} gives  the \textit{similarity equation}
\begin{equation}
\tnabla ^2 (-1+i\tpartial_1+(i/2)\tpartial_2) F = \tJ(F, \tnabla ^2 F),
\label{asympteuler}
\end{equation}
where the overscript tilde means that the partial derivatives are taken with 
respect to the new variables. The initial condition \rf{deuxmodes} becomes 
a boundary condition
\begin{equation}
F(\tz) \simeq \frac{1}{2} \left( \e ^{-i\tilde z_1} + \e ^{-2i\tilde z_2}
\right), \quad \tilde y_1\to -\infty,\,\,\,\,\tilde y_2\to -\infty.
\label{asymbound}
\end{equation}
Note that \rf{asympteuler} has no time variable. Its solution can be shown to
be analytic for sufficiently negative $\tilde y_1$ and $\tilde y_2$. If it has
singularities in the real or complex domain then, by \rf{deftz}, the solution
of the original Euler equation should have short-time singularities at a
distance $\delta(t) \propto \ln (1/t)$.\footnote{This law, for flow having
initially a finite number of Fourier modes, was first derived for the
one-dimensional Burgers equation and conjectured to apply also to the
multi-dimensional incompressible Euler equation \citep{frisch84}.}

The outline of the present paper is as follows. In Section~\ref{s:numshort}
we check the logarithmic law for $\delta(t)$ and thus the validity of the
similarity ansatz. In  Section~\ref{s:solving} we develop a new technique
for solving the similarity equation \rf{asympteuler} in both the real
and complex domains. In  Section~\ref{s:results} we present numerical
results on the nature and the geometry of the singularities. In  
Section~\ref{s:singular} we show how to actually construct the singular
manifold from the Fourier transform of the solution. In  
Section~\ref{s:conclusion} we make some concluding remarks.

\section{Numerical validation of the short-time behavior}
\label{s:numshort}

The standard way of measuring the distance $\delta$ of the nearest
complex-space singularity (also called the width of the analyticity strip) is
to use the method of tracing \citep{tracing}.  Indeed, the spatial Fourier transform
$\hat v_{\k}$ of a periodic function $v(\x)$ has its modulus decreasing at
large wavenumbers $k\equiv |\k|$ as $e ^{-\delta k}$ in both one and several
space dimensions. More precisely, for each direction $\hat \k\equiv \k/k$, 
the modulus of the
Fourier transform decreases as $e ^{-\delta_{\hat \k}k}$; the width of the
analyticity strip is then the minimum over all directions $\hat \k$ of
$\delta_{\hat \k}$ which -- by a steepest descent argument -- also controls the
high-$k$ decrease of the angle average of the modulus of the Fourier
transform.\footnote{For the way $\delta_{\hat \k}$ is related to the singular
manifold in the two-dimensional case, see Section~\ref{s:singular}.}

\begin{figure}[ht]
\iffigs
\centerline{\psfig{file=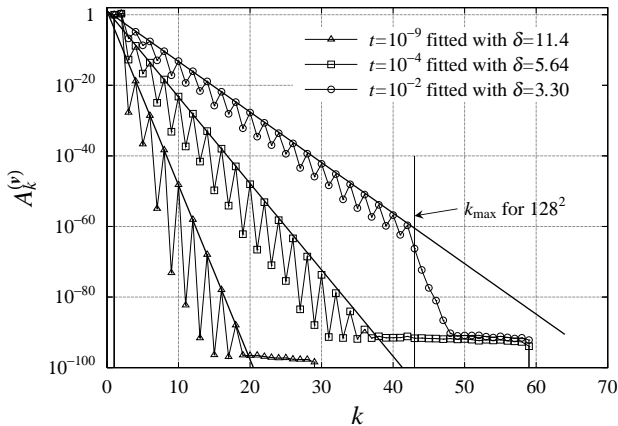,width=8.5cm,clip=}} 
\else\drawing 60 10 {short-time spectra (fig.6 report)}
\fi
\caption{Angular averages of the modulus of the velocity Fourier modes
at short times calculated with 90-digit precision. Strong oscillations
between odd and even $k$'s are due to the symmetry of the initial
condition \rf{deuxmodes}. The values of the time steps for
integrating the Euler equation with $64^2$ and $128^2$ grid points are
respectively $2\times10^{-11}$ and $5\times10^{-6}$.}
\label{f:shorttimespectra}
\end{figure}

\begin{figure}[ht]
\iffigs
\centerline{\psfig{file=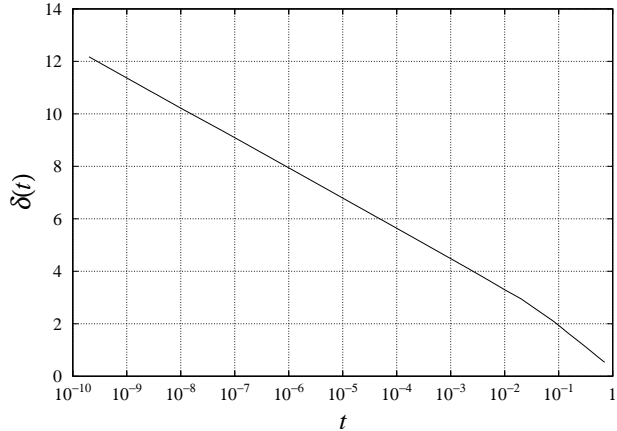,width=8.5cm,clip=}} 
\else\drawing 60 10 {logarithmic law}
\fi
\caption{Width of the analyticity strip $\delta(t)$ at short times
 $2\times10^{-10} \le t \le 7.3\times10^{-1}$ measured from the angular
 average of the modulus of the velocity Fourier modes restricted to even 
$k$'s.}
\label{f:deltat}
\end{figure}

Accurate measurement of the width of the analyticity strip $\delta$ gets
difficult when it becomes smaller than a few meshes, so that there is not
enough resolution to see long exponential tails \citep{brachetetal}. A
different difficulty appears when $\delta$ is very large and the Fourier
transform decreases so rapidly that it gets lost in roundoff noise at rather
small $k$'s.  To check on the logarithmic law of variation of $\delta(t)$ at
short times for the full time-dependent Euler equation \rf{streamform}, we
have to face the latter difficulty. To overcome it, we employed a 90-digit
multiprecision spectral calculation\footnote{All multiprecision calculations
in this paper were done using the package MPFUN90 \citep{bailey}.}  with the
number of grid points ranging from $64^2$ to $128^2$  The temporal scheme is
fourth order Runge-Kutta.  Fig.~\ref{f:shorttimespectra} shows the $k$
dependence at various short times of the angular averages\footnote{More
precisely, we use ``shell-summed averages'', defined at the beginning of
Section~\ref{s:results}.} of the modulus of the Fourier transform of the
velocity $\v \equiv (\partial_2 \psi,\, -\partial_1 \psi)$, where $\partial_1$
and $\partial_2$ are the derivatives with respect to $x_1$ and $x_2$. Note
that there is a strong odd-even wavenumber oscillation. This has to do with
the interference of two complex singularities separated by $\pi$ in the $x_2$
direction (a consequence of a symmetry of the initial condition).
A very clean exponential decrease is observed for even wavenumbers. This allows
the measurement of $\delta(t)$ for hundreds of values of $t$ covering 
a very wide range, as shown
in Fig.~\ref{f:deltat} in log-lin coordinates.
It is seen that, up to approximately $t=10^{-2}$, the logarithmic law is
satisfied over eight decades, a range which could not be achieved
without the use of very high-precision computations.

\section{Solving the similarity equation}
\label{s:solving}

\begin{figure}[ht]
\iffigs
\centerline{\psfig{file=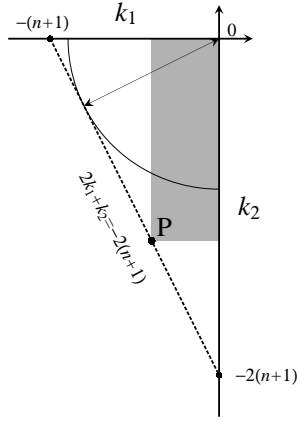,width=4cm,clip=}}
\else\drawing 60 10 {quadrant}
\fi
\caption{Region in the wavenumber space relevant in calculating the Fourier
 coefficients $\hat{F}_{k_1,\, k_2}$ on the line $2k_1 + k_2 = -2(n + 1)$.
 The largest disk centered at the origin fitting into this region is also
 shown.}
\label{f:quadrant}
\end{figure}

We are interested in space-periodic solutions to \rf{asympteuler} satisfying
the boundary condition \rf{asymbound}. This problem can be solved exactly
using Fourier series
\begin{equation}
F(\z) = \sum_{k_1=-\infty}^{\infty}\sum_{k_2=-\infty}^{\infty} \hat
F_{k_1,k_2} e ^{i(k_1z_1+k_2z_2)}.
\label{fourier}
\end{equation}
Note that we have dropped all tildes on the space variables since from now on
we will work exclusively with the similarity variables.  Obviously, the
boundary condition \rf{asymbound} allows the presence only of Fourier
harmonics with wavevectors in the quadrant $(k_1\le 0,\,\, k_2\le
0)$. After \rf{asympteuler} is rewritten in terms of the Fourier coefficients
of $F$, the multiplications appearing in the Jacobian go over into simple
convolutions with only finitely many terms, because all wavevectors involved
must have non-positive components. It follows that, for integer $n\ge 1$, all
Fourier coefficients for wavevectors which are on the line $2k_1+k_2 =-2(n+1)$
(see Fig.~\ref{f:quadrant})
are expressible in terms of the finitely many Fourier coefficients with 
wavevectors lying  above this  line. 
More precisely, for any point P on this line, the region of dependence
is the grey-shaded rectangle. 
Specifically,
we define the selective  Fourier sum over  this line 
\begin{equation}
 F_n(\z) \equiv \sum_{\sigma = 0}^{n + 1} \hat F_{-\sigma,-2(n + 1 - \sigma)}
 e ^{-i(\sigma z_1 + 2(n + 1 - \sigma)z_2)},
 \label{e:dominant}
\end{equation}
so that
\begin{equation}
F(\z) = \sum_{n=0}^\infty F_n(\z).
\label{}
\end{equation}
From \rf{asympteuler} we then obtain the recursion relations
\begin{equation}
 \nabla^2 F_{n + 1}
  =
  \frac{1}{n + 1}
  \sum_{r= 0}^n
  J(F_r, \nabla^2F_{n-r}),
\label{e:psi_n_recursion}  
\end{equation}
with
\begin{equation}
F_0(\z) =\frac{1}{2}\left( e ^{-iz_1} + e ^{-2iz_2}\right).
\label{initF}
\end{equation}
Equation \rf{e:psi_n_recursion} can then be rewritten as a set
of algebraic equations for the Fourier coefficients:
\begin{eqnarray}
\hat{F}_{k_1,\,k_2} &=& 
\sum_{p_1 = k_1}^{0}
\sum_{p_2 = k_2}^{0}
T_{\k}({\bm p}) \,
\hat{F}_{p_1,\,p_2} \, 
\hat{F}_{k_1 - p_1,\,k_2 - p_2} \,,
\label{e:beta_recursion}\\
T_{\k}({\bm p}) &\equiv& 
\frac{2}{2k_1 + k_2 +2} 
\frac{|\k - {\bm p}|^2 ({\bm p} \wedge \k)}{|\k|^2} \,,
\label{defT}
\end{eqnarray}
where $\k\wedge \k'\equiv k_1k'_2 -k_2k'_1$.
One could solve the Poisson equations
\rf{e:psi_n_recursion} which recursively define the $F_n$'s using
FFT methods as in \citet{brachetetal}.  Alternatively -- and this is the
method used here -- equations~\rf{e:beta_recursion}-\rf{defT} can be used to calculate recursively
the {\it exact\/} expressions of all the Fourier coefficients.
The first two $F_n$'s are
\begin{eqnarray}
F_1(\z) &=& -\frac{3}{10} e ^{-i(z_1+2z_2)},\\
F_2(\z) &=& -\frac{3}{340} e ^{-i(z_1+4z_2)}
             +\frac{3}{40} e ^{-i(2z_1+2z_2)}.
\label{F1,F2}   
\end{eqnarray}

We have numerically determined the $F_n$'s for $n$ up to $n_{\rm max} = 1500$,
using quadruple-precision (35-digit) accuracy. As we shall see in the next
section, lower accuracy would give spurious results. From
\rf{e:beta_recursion} and \rf{defT}, the $\hat{F}_{k_1,\,k_2}$ are
obviously real.  Furthermore, we find numerically that their signs alternate:
specifically $(-1)^{k_1}\hat{F}_{k_1,\,k_2} \ge 0$, for all $(k_1,k_2)$
except $(-1,0)$.

\section{Numerical results on singularities}
\label{s:results}

We shall work here mostly with shell-summed (Fourier) amplitudes. For a 
given periodic function $f$, we define its shell-summed amplitude as
\begin{equation}
A^{(f)}_K \equiv \sum_{K\le |\k| < K+1} |\hat f_\k|,
\label{defssa}
\end{equation}
where the $\hat f_\k$ are the Fourier coefficients of $f$. The shell-summed
amplitudes for the solution $F$ to the similarity equation \rf{asympteuler}
can be calculated, in principle exactly, for all wavenumbers $K$
such that $K+1 < k_{\rm max}$ where 
$k_{\rm max} = (2/\sqrt 5)(n_{\rm max}+1)$, this being the radius
of the largest disk centered at the origin and contained in the 
triangular region of Fig.~\ref{f:quadrant}.

\begin{figure}[ht]
\iffigs
\centerline{%
\psfig{file=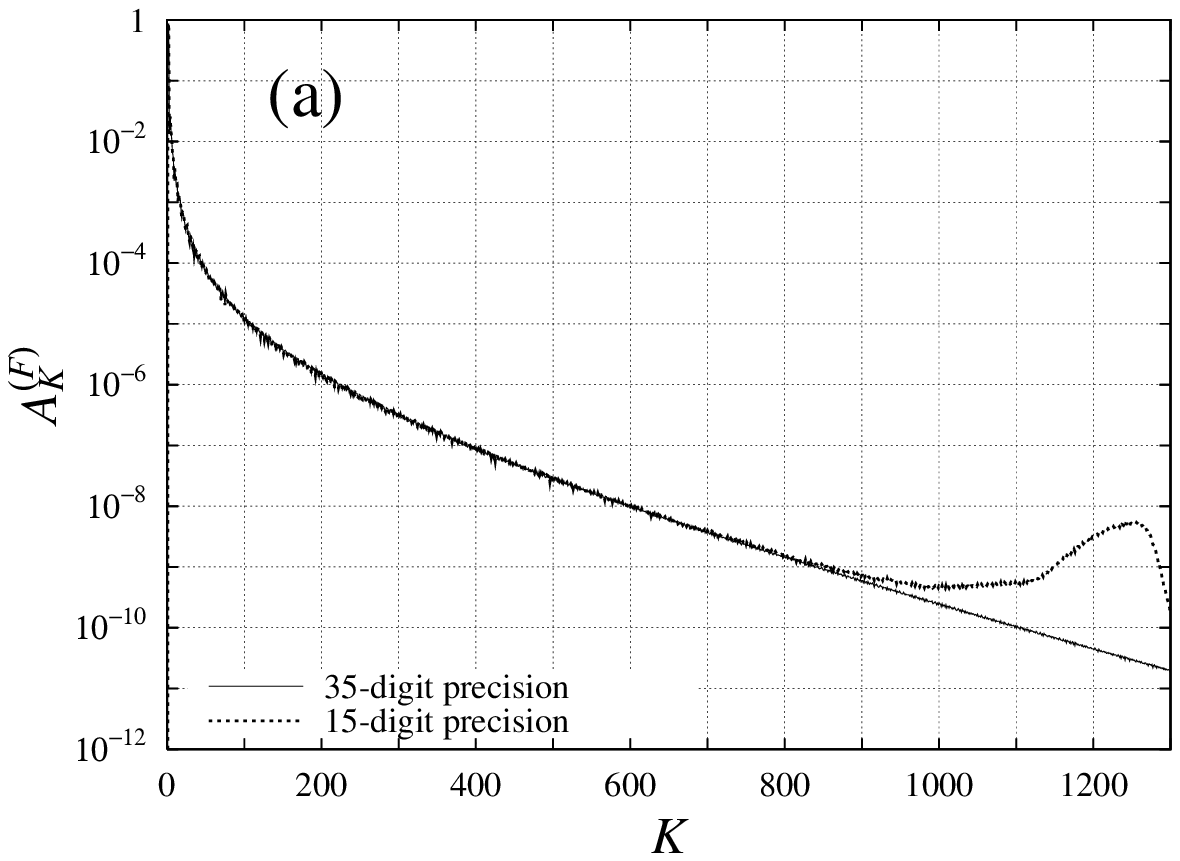,width=8.5cm,clip=}%
\psfig{file=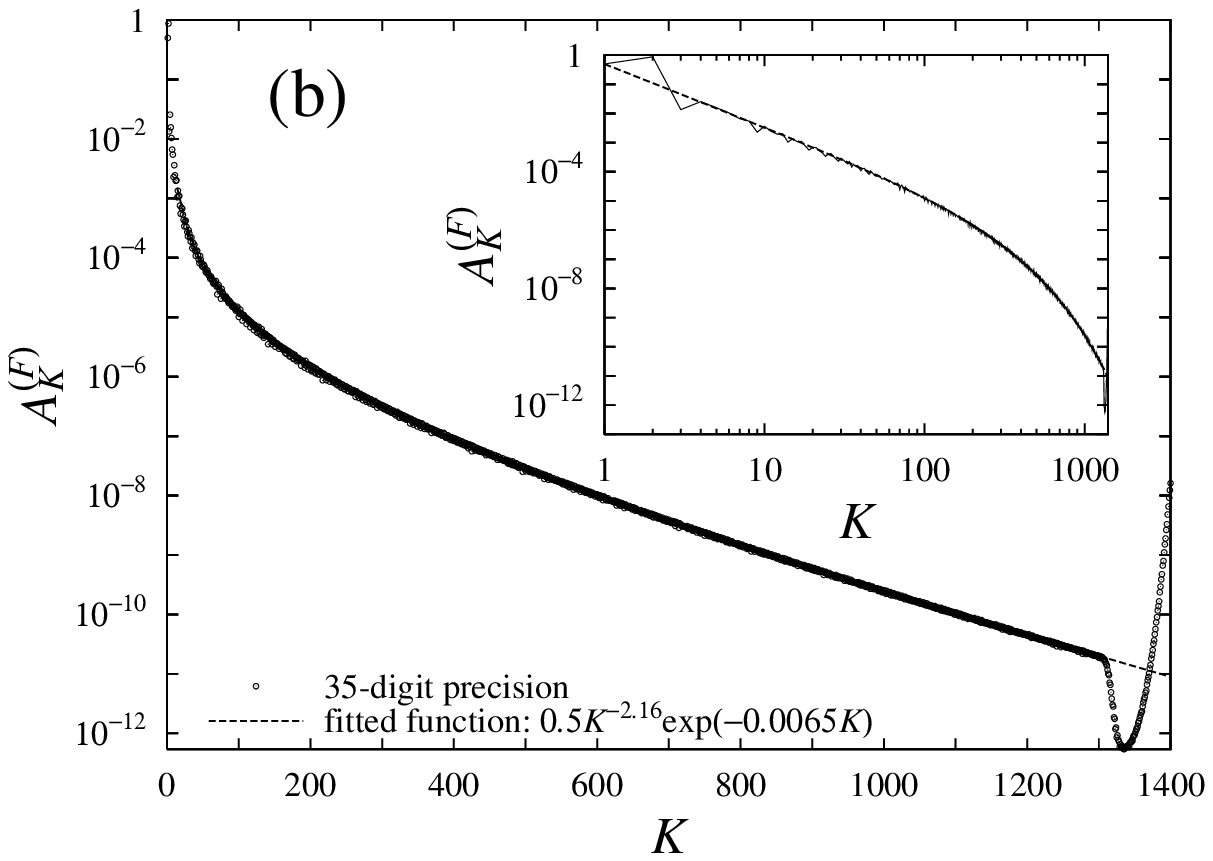,width=8.5cm,clip=}%
}
\else\drawing 60 10 {ssa for F with loglog as inset}
\fi
\caption{(a) Shell summed amplitudes $A^{(F)}_K$ for the stream function $F$
 calculated with 15 and 35-digit precisions in lin-log coordinates.
 The strange behavior of the 15-digit result in the high wavenumber
 region $k \ge 800$ is due to insufficient precision.
 (b) Shell summed amplitude $A^{(F)}_K$ calculated with 35-digit
 precision and almost overlapping least square fit to $C K^{-\alpha} \exp
 (-\delta K)$ in lin-log coordinates.
 Numerical instability is observed for $K > 1300$.
 Inset: $A^{(F)}_K$ with 35-digit precision and its least square fit
 in log-log coordinates.}
\label{f:ssaF}
\end{figure}

Fig.~\ref{f:ssaF} shows, as a function of $K$, the shell-summed amplitude of
$F$ for $n_{\rm max} =1500$. As seen in Fig.~\ref{f:ssaF}(a), the results
obtained with 15 and 35 digit precisions differ markedly beyond wavenumber
800.  Even the 35-digit calculation becomes unreliable beyond wavenumber
1300. In FMB, when we discussed results about singularities without resorting
to short-time asymptotics, we reported various difficulties: the need to
perform a kind of Krasny filtering \citep{krasny, majda-bertozzi, caflishetal}
and our failure to improve the range of scaling by going to resolutions higher
than $512^2$. It is now clear that the key to clean results is to use high
precision.

We fit the shell-summed amplitude of $F$ by a function of the form $C
K^{-\alpha} \exp (-\delta K)$ and find the best fit to be $0.5 K^{-2.16}
\exp(-0.0065 K)$, as shown in Fig.~\ref{f:ssaF}(b). The fit is done by least
squares in the lin-log representation over the interval $0 \le K \le 1100$.

Hence, the solution to the similarity equation has complex-space
singularities, the closest one being within $\delta \approx 0.0065$.  This
relatively small value, for an equation in which all the coefficients are
order unity, is accidental and can be changed by slightly modifying the
coefficients  in front of the two harmonics in the initial condition
\rf{deuxmodes} and in \rf{initF}, as will be shown later in this section.

\begin{figure}[ht]
\iffigs
\centerline{\psfig{file=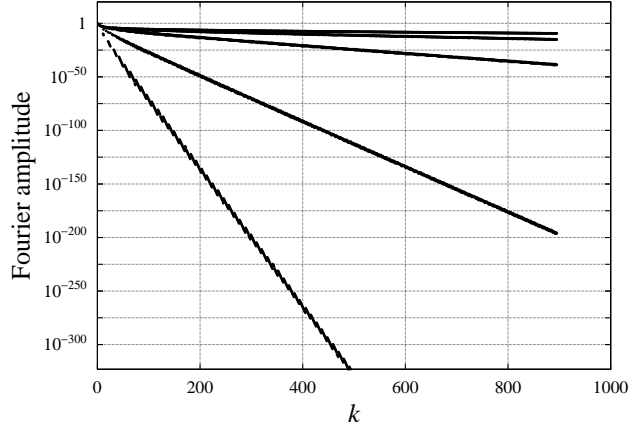,width=8.5cm,clip=}} 
\else\drawing 60 10 {lin-log of ampl. for various theta}
\fi
\caption{Amplitudes of the Fourier modes $|\hat{F}_{\k}|$ along five
 different directions from $1.05\pi$ to $1.32\pi$ (from bottom to top)}
\label{f:thetadepampl}
\end{figure}

\begin{figure}[ht]
\iffigs
\centerline{\psfig{file=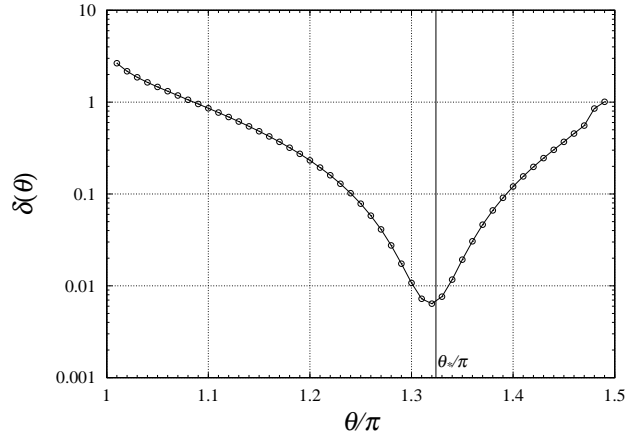,width=8.5cm,clip=}} 
\else\drawing 60 10 {deltaoftheta}
\fi
\caption{Logarithmic decrement $\delta(\theta)$ as a function of
 the polar angle $\theta$ in the $(k_1, \, k_2)$ plane. The minimum of
 $\delta(\theta)$ is around $0.0065$.
 The most singular direction is $\theta_\star = 1.324\pi$.
 }
\label{f:deltaoftheta}
\end{figure}

\begin{figure}[ht]
\iffigs
\centerline{\psfig{file=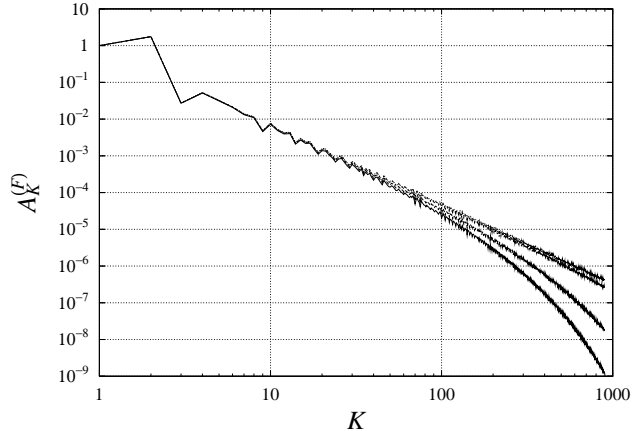,width=8.5cm,clip=}} 
\else\drawing 60 10 {loglog of shifted}
\fi
\caption{Shell summed amplitudes of the 
 stream functions $F^{(h)}$
 shifted by $(ih\cos \theta_\star, \, ih\sin \theta_\star)$ into the
 most singular direction $\theta_\star$, for $h = 0.0, 0.0030, 00060,
 0.0065$ (bottom to top). }
\label{f:ssashifted}
\end{figure}

The angular dependence of the mode amplitude at high wavenumbers can also be
obtained. Fig.~\ref{f:thetadepampl} shows that $|\hat F_\k|$ decreases
exponentially with $k \equiv |\k|$ for a given direction $\hat \k =(\cos
\theta,\,\sin\theta)$.\footnote{Actually, we sum over all Fourier modes for
which $\hat \k$ is within $\pi/100$ of the direction $\theta$.}  We see that
the logarithmic decrement, denoted by $\delta(\theta)$, varies strongly with
$\theta$. Its variation is shown in Fig.~\ref{f:deltaoftheta} 
for $\pi<\theta < 3\pi/2$, that is over the third angular quadrant where
$k_1$ and $k_2$ are negative.\footnote{At the edges of this
quadrant $\delta(\theta)$ becomes infinite (W.~Pauls, private
communication), but this is hidden by the slight angular averaging.} 
Obviously, the
width of the analyticity strip is
\begin{equation}
\delta = \min_\theta \, \delta(\theta),
\label{deltamin}
\end{equation}
the minimum being achieved in the \textit{most singular direction} $\theta
=\theta_\star$, such that $\tan \theta_\star \approx 1.62$.  The function
$\delta(\theta)$ can be related to the singular manifold (see Section~\ref{s:singular}).
As in one dimension, the prefactor of the exponential in the shell-averaged
amplitude $A_K^{(F)}$ contains information about the nature of the
singularities. In the inset of Fig.~\ref{f:ssaF} we see about one decade of
power-law scaling before the exponential falloff.  This range can be increased
by moving closer to the singularity, more precisely to the point $\z_\star$ on
the singular manifold closest to the real domain which has imaginary part
${\rm Im}\, \z_\star = (y_{1\star},\,y_{2\star})$ with $y_{1\star} = \delta
\cos \theta_\star$ and $y_{2\star} = \delta \sin \theta_\star$.  Such an
imaginary shift produces a function $F^{(h)}$ whose Fourier coefficients are
obtained by multiplying the Fourier coefficients of $F$ by $\exp\left(hk_1\cos
\theta_\star +hk_2\sin \theta_\star \right)$, where
$0<h<\delta$.\footnote{Note that, in the short-time asymptotics, an imaginary
shift $(h_1,\, h_2)$ amounts to changing the initial condition \rf{deuxmodes}
into $\psi_0(\z) = e^{h_1}\cos (z_1) + e^{2h_2}\cos (2z_2)$ within terms
irrelevant for $y_1\to +\infty$ and $y_2 \to +\infty$.}
Fig.~\ref{f:ssashifted} shows shell-summed amplitudes of $F^{(h)}$ for four
values of $h$. For $h=\delta$ more than two decades of power-law scaling is
seen with an exponent somewhere between $-2.0$ and $-2.2$. This clean scaling
evidence has an important consequence for the vorticity on the singular
manifold (see below).

\begin{figure}[ht]
\iffigs
\centerline{%
\psfig{file=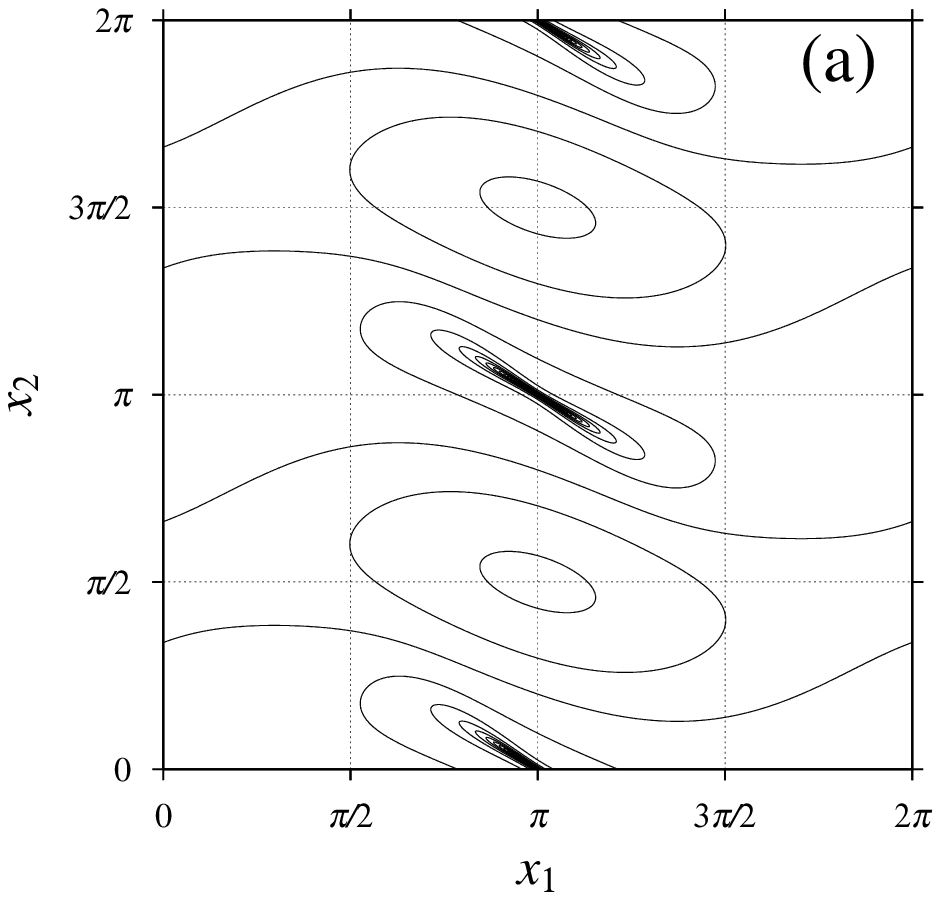,width=8.0cm,clip=}%
\psfig{file=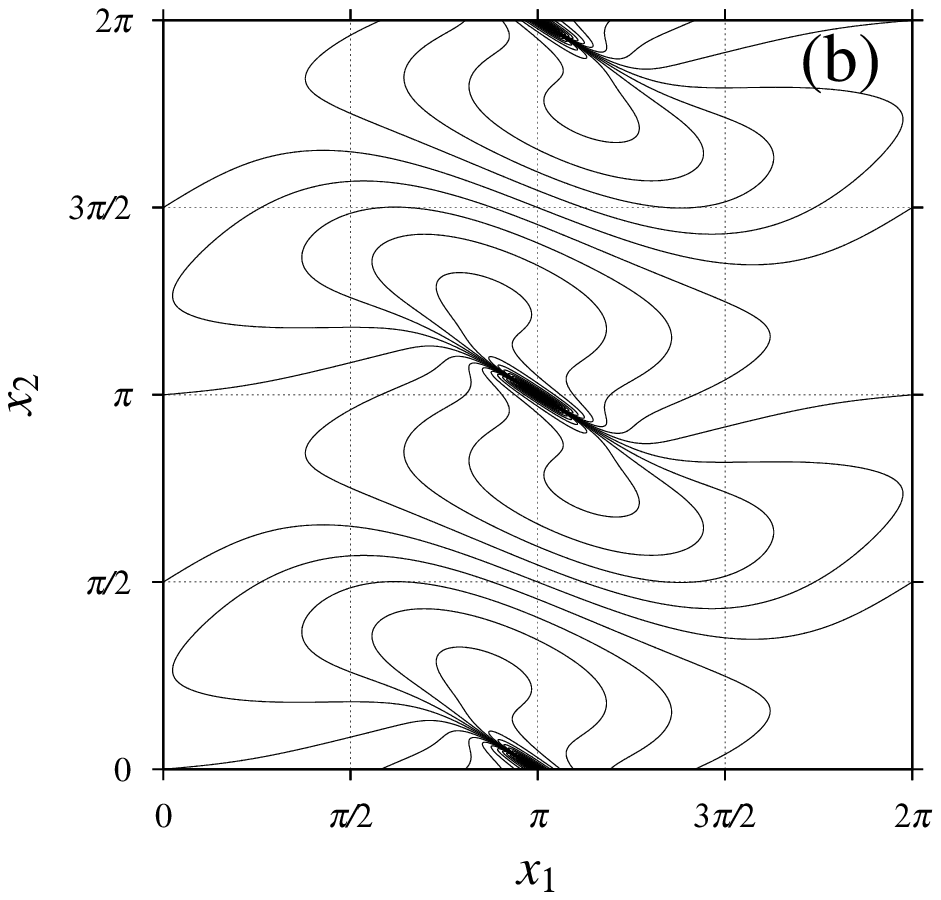,width=8.0cm,clip=}%
}

\vspace{5mm} 
 
\centerline{%
\psfig{file=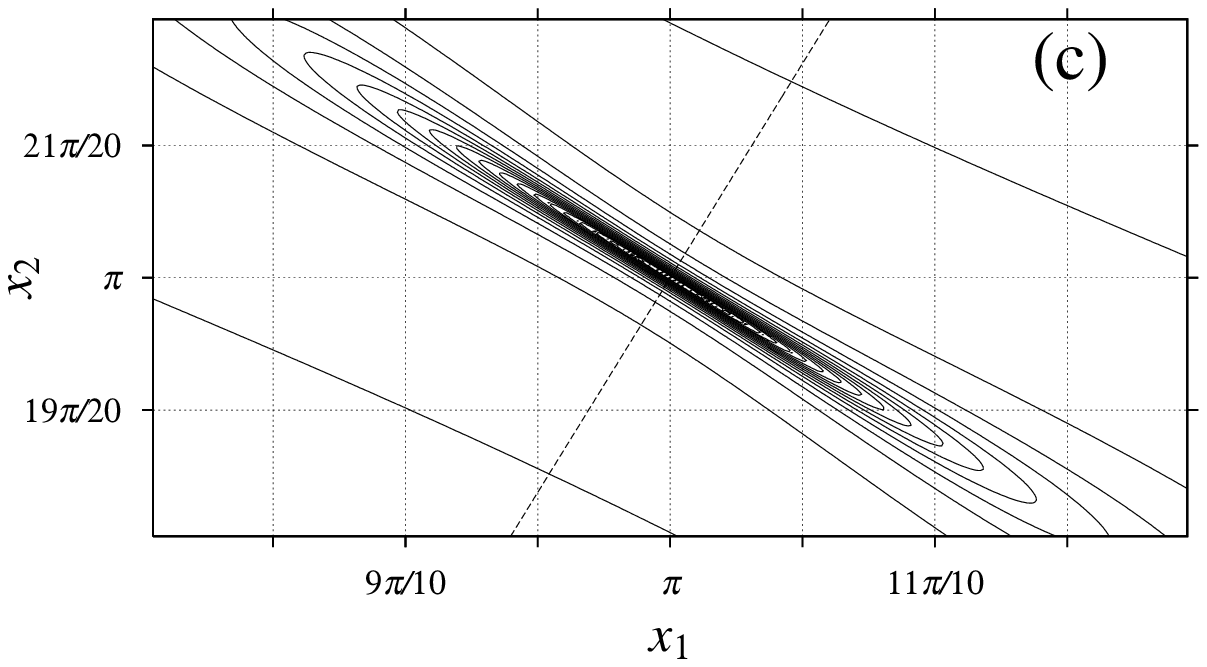,width=8.5cm,clip=}%
\psfig{file=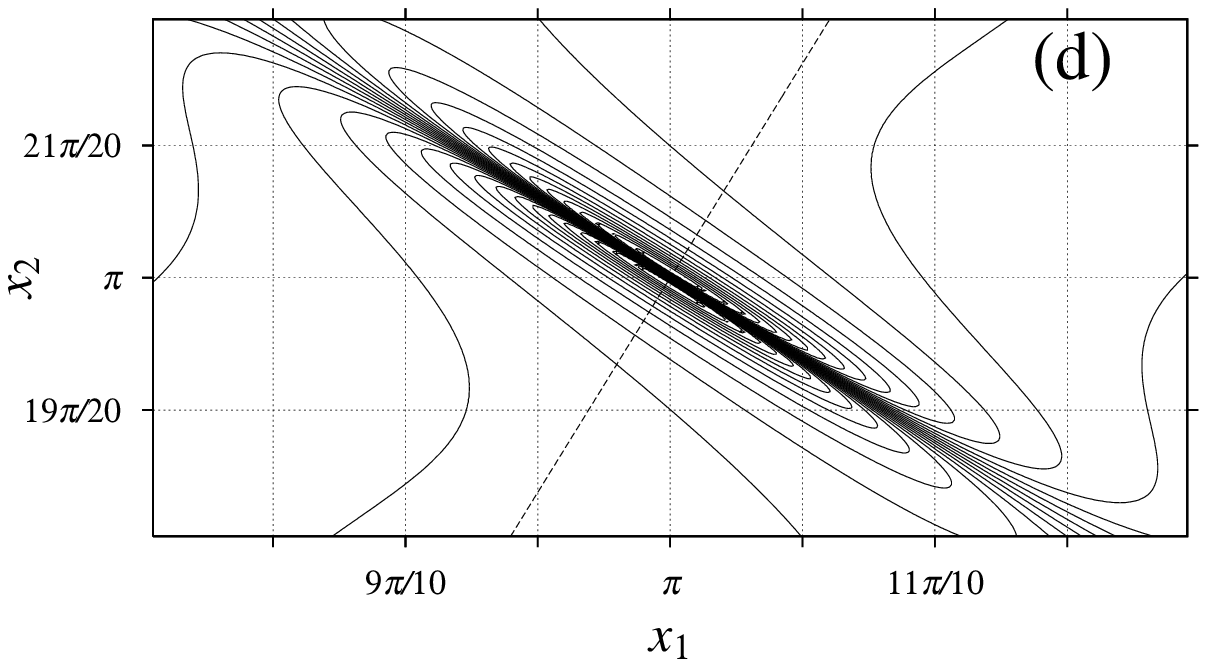,width=8.5cm,clip=}%
} 
\else\drawing 60 10 {}
\fi
\caption{%
Contours of the vorticity, (a) real part, (b) imaginary part,
 (c) and (d) enlargements of (a) and (b).
} 
\label{f:contours}
\end{figure}

\begin{figure}[ht]
\iffigs
\centerline{%
\psfig{file=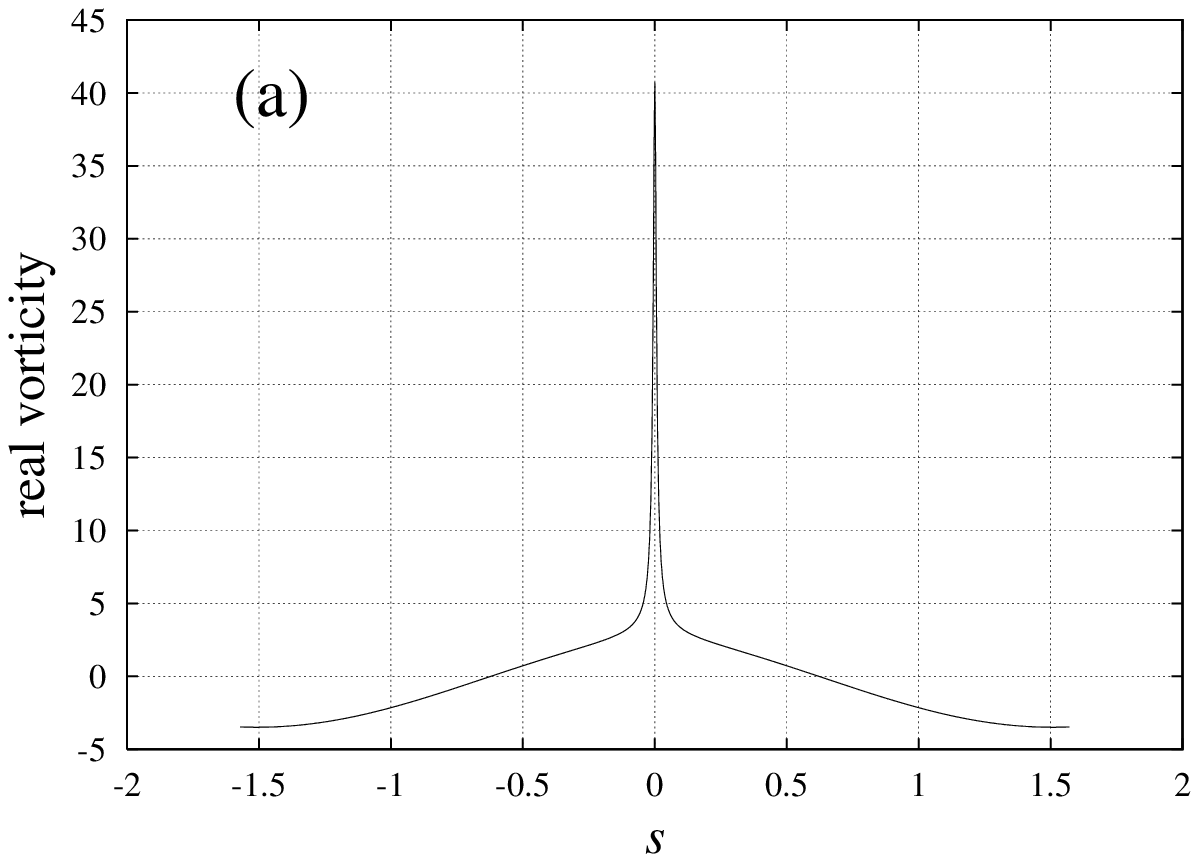,width=8.5cm,clip=} 
\psfig{file=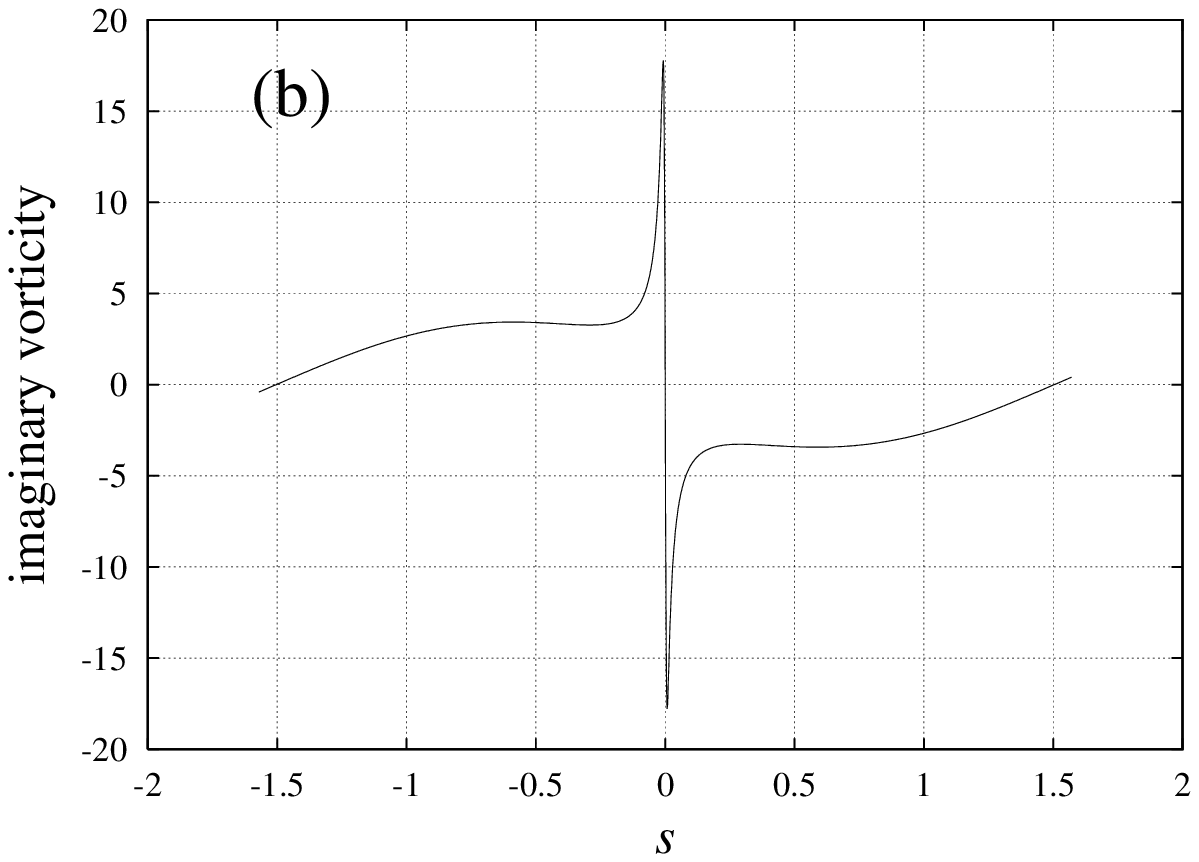,width=8.5cm,clip=} 
} 
\centerline{%
\psfig{file=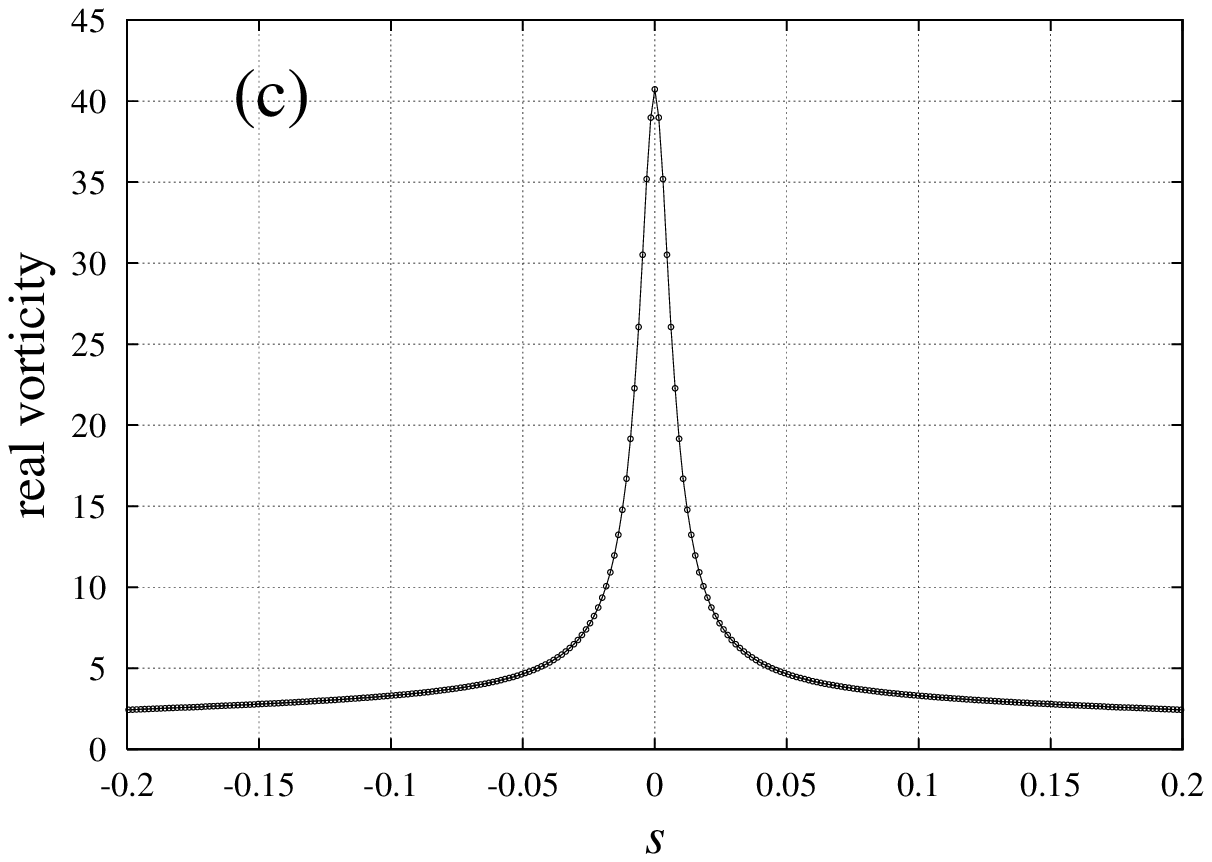,width=8.5cm,clip=}
\psfig{file=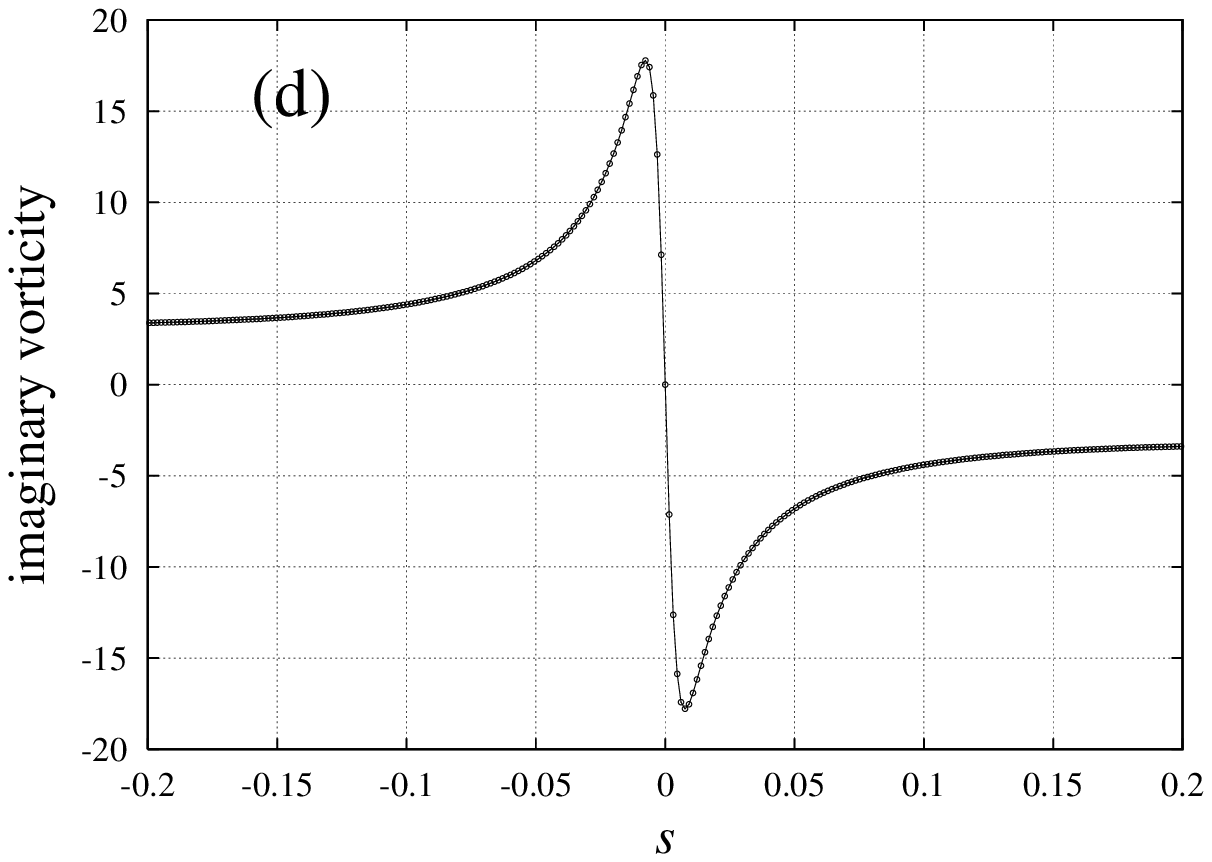,width=8.5cm,clip=}
} 
\else\drawing 60 10 {4 cuts in the most singular direction}
\fi
\caption{%
 Variation of the vorticity along a cut through $(x_1, \, x_2)=(\pi, \,
 \pi)$ with the most singular direction $\theta_\star$.
(a) real part, (b) imaginary part, (c) and (d) enlargements of (a) and (b).}
\label{f:singcuts}
\end{figure}

\begin{figure}[ht]
\iffigs
\centerline{\psfig{file=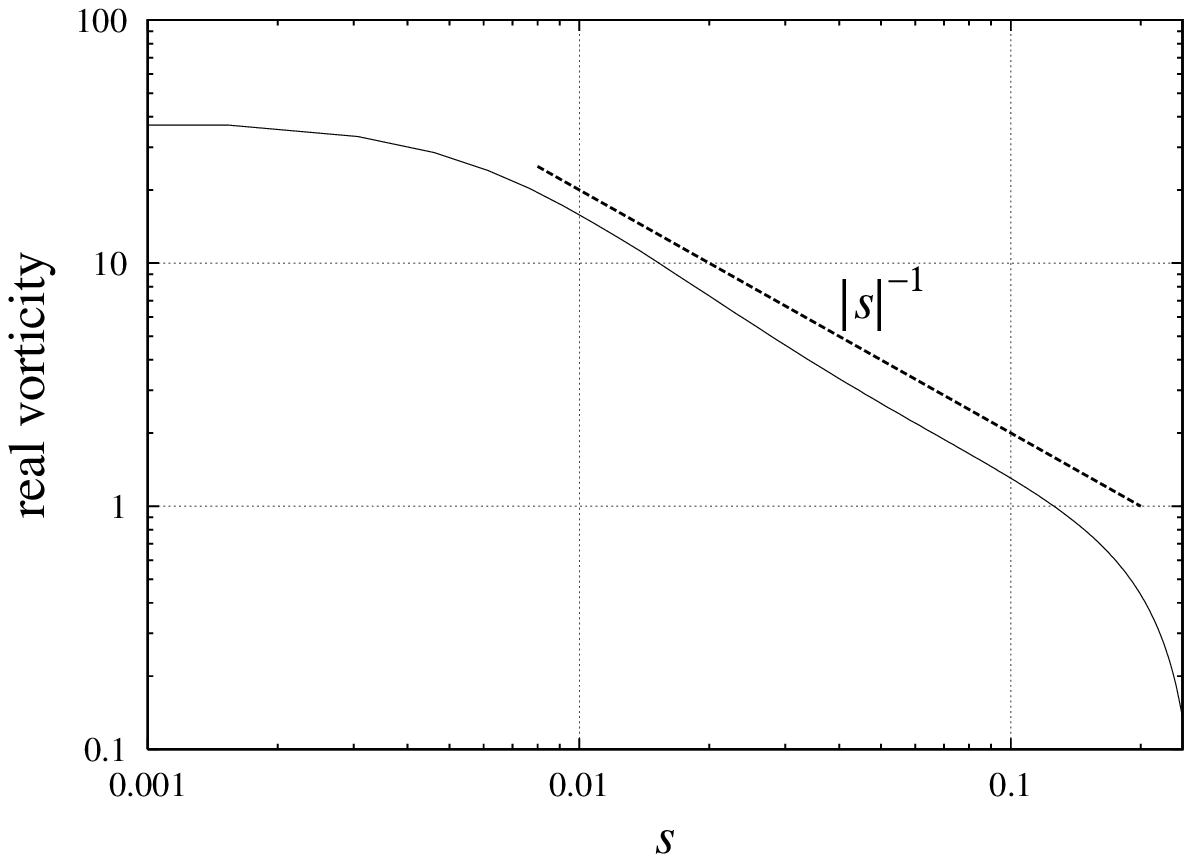,width=8.5cm,clip=}} 
\else\drawing 60 10 {1overx fit for vorticity}
\fi
\caption{
 Same data as in the right half of Fig.~\ref{f:singcuts} (c) in log-log
 coordinates. To obtain a better scaling, we subtract a constant (here
 $2.0$) from the real part of the vorticity.
 }
\label{f:1overxfit}
\end{figure}

To gain additional insight, we now show results in physical space in terms
of the real and imaginary parts of the vorticity $\omega \equiv -\nabla ^2
F$.\footnote{Since the Fourier transform is real and supported in the quadrant
$(k_1\le 0,\, k_2\le 0)$ there are actually Kramers--Kronig relations between
the real and imaginary parts.} For this we limit ourselves to $n_{\rm
max}=1000$. As a consequence, $k_1$ runs from $-1001$ to 0, while $k_2$ varies
from $-2002$ to 0.  We use an FFT program with $4096^2$ grid
points. Fig.~\ref{f:contours} shows contours of the real and imaginary parts
of the vorticity. The symmetries seen are a consequence of dynamically
preserved symmetries in the initial condition \rf{deuxmodes}. Near the center
$x_1=\pi,\,\, x_2=\pi$ there is a highly anisotropic large-amplitude
sheet-like structure which can be interpreted as the manifestation of a smooth
singular manifold which gets within roughly four meshes of the real
domain.\footnote{This manifold will actually be constructed explicitly in
Section~\ref{s:singular}.} By far, the fastest vorticity variation is obtained
perpendicularly to the sheet-like structure, in the most singular direction
$\theta_\star$. Fig.~\ref{f:singcuts} shows the variation of the vorticity
along a cut through $x_1=\pi,\,\, x_2=\pi$ in the most singular direction. It
is seen that the vorticity becomes rather large (around 40 for the real
part). The behavior of the real part near the peak, as a function of the
distance $s$ to the peak is very roughly as $1/|s|$ as seen in
Fig.~\ref{f:1overxfit}.

\begin{figure}[ht]
\iffigs
\centerline{\psfig{file=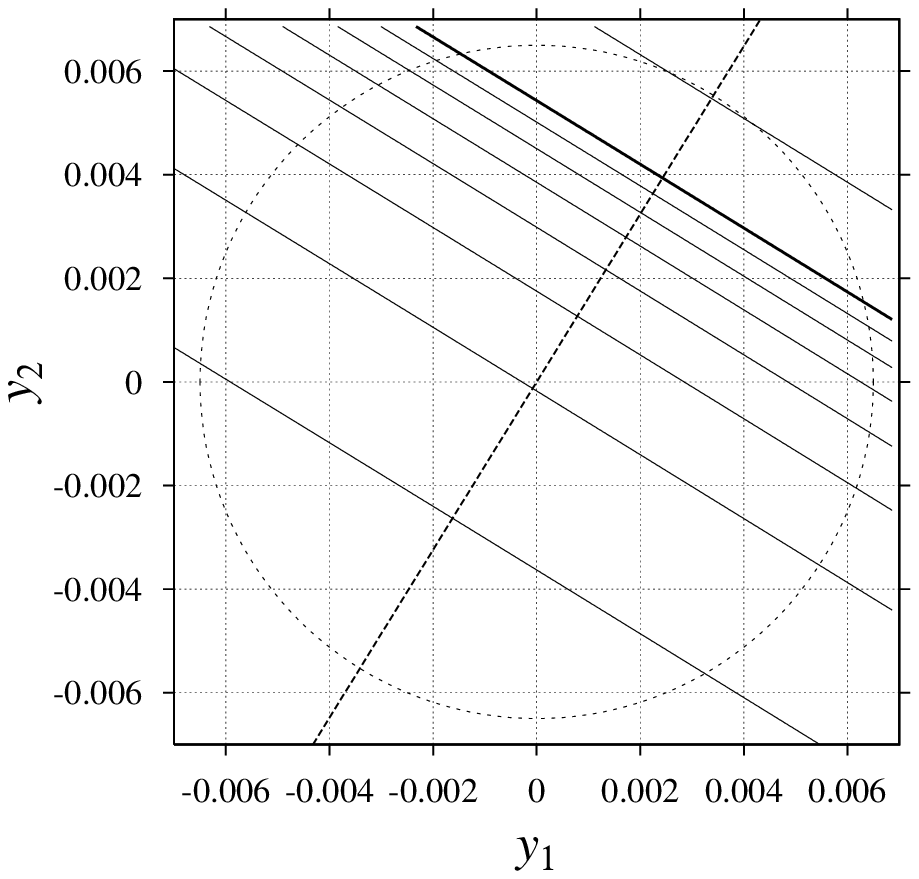,width=10.0cm,clip=}} 
\else\drawing 60 10 {straight}
\fi
\caption{
 Contours of the vorticity $\omega_\star(y_1, \, y_2)$ for $x_1 =
 \pi$ and $x_2 = \pi$. The dotted circle has radius $\delta = 0.0065$,
 whose smallness makes the contours nearly straight.
 Contour values from lower left to upper right are 
 $30, 40, \ldots, 90, 100, 200$ (the thick solid line is for $100$).}
\label{f:straight}
\end{figure}

\begin{figure}[ht]
\iffigs
\centerline{\psfig{file=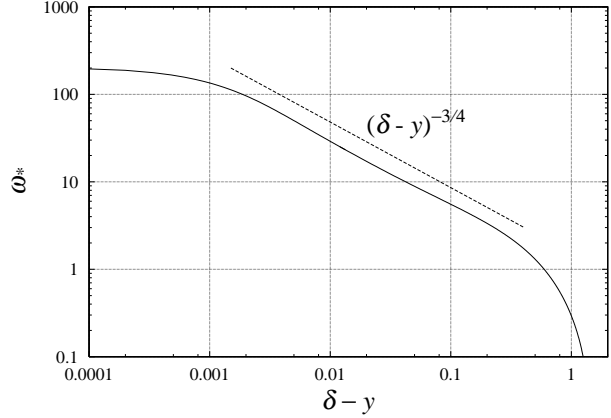,width=8.5cm,clip=}} 
\else\drawing 60 10 {threequarters}
\fi
\caption{Vorticity $\omega_\star$ along the most singular direction as a
 function of the distance $\delta - y$ to the singularity nearest to the
 real domain.}
\label{f:threequarters}
\end{figure}

It is also of interest to show the variation of the vorticity in the 
$(y_1,\,y_2)$-plane, that is $\omega_\star(y_1,y_2) \equiv \omega (\pi+iy_1,\pi+iy_2)$. Symmetry
implies that this is a real quantity which, by \rf{fourier},
can be written as
\begin{eqnarray}
\!\!\!\!\!\!\omega_\star (y_1,y_2)&=&
  \sum_{k_1 = -\infty}^{0}
  \sum_{k_2 = -\infty}^{0}
  (-1)^{k_1} k^2\, \hat{F}_{k_1,\,k_2}
  \, e^{\, -(k_1 y_1 + k_2  y_2) \, },
\label{e:omega_pipi}\\[1ex]
k^2&=&   k_1^2 + k_2^2. \nonumber
\end{eqnarray}
The latter can be viewed as a double power series in the variables $w_1\equiv
e ^{y_1}$ and  $w_2 \equiv e ^{y_2}$.
The alternating sign property of the $\hat{F}_{k_1,\,k_2}$ mentioned
at the end of  Section~\ref{s:solving} implies that all the coefficients
$(-1)^{k_1} k^2\, \hat{F}_{k_1,\,k_2}$ are nonnegative for $k_1 < -1$. It is well
known that if a power series in one variable with nonnegative coefficients 
defines an analytic function with singularities, then the nearest singularity
to the origin is on the real positive axis 
\citep[Vivanti's theorem, see, e.g., ][]{vivanti}. 
Here we expect that there will be a whole singular curve
in the $(y_1,\,y_2)$-plane. The Fourier series can be used
to calculate $\omega_\star(y_1,y_2)$, at least within the disk
$y_1^2+y_2^2 <\delta ^2$. Fig.~\ref{f:straight} shows contours
of $\omega_\star$ in and around this disk. The contours are almost straight
lines perpendicular to the most singular direction, because $\delta$ is very
small. The variation of $\omega_\star$ along the the most singular direction
as a function of the distance $y$ to the origin is shown in
Fig.~\ref{f:threequarters}. When plotted as a function of the distance $\delta
-y$ to the nearest singularity $(y_{1\star},\,y_{2\star}) = \delta(\cos
\theta_\star,\, \sin \theta_\star)$, it follows roughly a power law with an
exponent close to $-3/4$ over almost two decades. We observe that the
vorticity reaches values around 200. Peak values between 100 and 200
and a scaling exponent close to $-3/4$ are also obtained for the behavior
of $\omega_\star$ near other points of the singular manifold (to be defined
precisely in the next section).
We observe that there may be an inconsistency between the exponent
observed in the real direction which is roughly $-1$ 
(Fig.~\ref{f:1overxfit}) and the exponent $-3/4$ observed in the
imaginary direction (Fig.~\ref{f:threequarters}) since the
vorticity, being analytic, should have the same scaling. The
singularities being contained in the imaginary plane above 
$x_1=\pi,\,x_2=\pi$, at least a few meshes away from the real domain,
the scaling in the imaginary direction is expected to be more reliable. 

Actually, there is strong evidence for the vorticity being infinite at
$(y_{1\star},\,y_{2\star})$ coming from the behavior of the shell-summed
amplitudes. Indeed, we know (i) that all the terms in the double sum
\rf{e:omega_pipi} are nonnegative and (ii) that the shell-summed stream
function amplitude decreases nearly as $k^{-2}$ over two decades or more
(Fig.~\ref{f:ssashifted}). It follows that the shell-sums for the vorticity at
$(y_{1\star},\,y_{2\star})$ are almost $k$-independent and thus, when we sum
over $k$, we get an infinite value.

\section {The singular manifold}
\label{s:singular}

In one dimension it is well known that the high-wavenumber asymptotics of the
Fourier transform of an analytic periodic function is governed
by the singularities nearest to the real domain 
\citep[see, e.g., ][]{carrier-krook-pearson, frisch-morf}. 
This has a non-trivial extension
to more than one dimension when the function has singularities on a 
complex manifold ${\mathcal S}$ determined by an equation $S(\z) =0$; let 
us sketch this in the  two-dimensional case.
Consider a periodic analytic function $F(\z)$ given
by the Fourier series \rf{fourier}. The Fourier coefficients
are given by the double integral over the real periodicity domain
\begin{equation}
 \hat F_{k_1,k_2} =\frac{1}{(2\pi)^2} \int\int dx_1dx_2 \, e 
 ^{-i(k_1x_1+k_2x_2)} F(\x).
 \label{invfourier}
 \end{equation}
 Let us set $k_1=k \cos \theta$ and $k_2= k\sin \theta$. We are 
 interested in the behavior of $\hat F_{k_1,k_2}$ when $k\to \infty$ for
 a given direction defined by the angle $\theta$. For this we change 
 coordinates from $(x_1,\, x_2)$ to $(x_\|,\, x_\perp)$ chosen, respectively, 
 parallel and perpendicular to $\k$. Note that the exponential
 in \rf{invfourier} involves only $x_\|$. Hence \rf{invfourier} can
 be rewritten as a one-dimensional Fourier transform
 \begin{eqnarray}
     \hat F &=&\frac{1}{2\pi}\int dx_\|\, e 
     ^{-ikx_\|}g_\theta(x_\|)
     \label{onedim}\\
    g_\theta(x_\|)&\equiv&  \frac{1}{2\pi}\int dx_\perp \, 
    F(x_\|\hat \k +x_\perp \hat \k_\perp),
 \label{intxperp}
 \end{eqnarray}
where $\hat \k \equiv \k/k$ and $\hat \k_\perp$ is $\hat \k$ rotated by
$+\pi/2$. It follows from \rf{onedim} that, for $k\to\infty$, $\hat F \sim e
^{-i kz^\star_\|}$ where $z^\star_\|$ is the singularity of $g_\theta(z_\|)$ 
in the complex $z_\|$
plane nearest to the real domain and with negative imaginary
part.\footnote{For simplicity, we are ignoring algebraic prefactors which
depend on the nature of the singularity.} Hence $|\hat F| \sim e
^{-k\delta(\theta)}$ where $\delta(\theta) = -{\rm Im}\,z^\star_\|$.

How do we obtain the singularities of $g_\theta(z_\|)$, given by the integral
\rf{intxperp}? For some (complex) $z_\|$ there may be singularities of
$F(z_\|\hat \k +z_\perp \hat \k_\perp)$ for real $z_\perp$. They can however
be avoided by shifting the contour of integration away from the real $z_\perp$
axis.  If we change $z_\|$, this will work as long as the contour does not get pinched between two (or
more) coalescing singularities (the pinching, generically, does not take place
on the real $z_\perp$ axis). Assuming that the singular manifold ${\mathcal
S}$ can be represented by $S_\theta(z_\|,z_\perp)=0$ in the $z_\|$ and
$z_\perp$ variables, a necessary condition for this pinching is a double root
in $z_\perp$, i.e. $S_\theta=0$ and $\partial S_\theta /\partial z_\perp
=0$. This system of two equations has discrete solutions
$(z^\star_\|,\,z^\star_\perp)$, one of which will control the
asymptotics.
In an earlier version of this paper (available at 
http://arxiv.org/pdf/nlin.CD/0310044v1) we stated  that the
pinching argument may already be known. Recently W.~Pauls pointed out
to us that it has already
been used in a special case by Henri Poincar\'{e} 
\citep[see Sections 94--96 of][]{poincare} and has later
been generalized by \citet{tsikh}. 

\begin{figure}[ht]
\iffigs
\centerline{\psfig{file=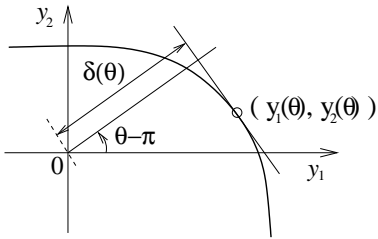,width=5.0cm,clip=}} 
\else\drawing 60 10 {delta of theta from sing. manifold}
\fi
\caption{Geometrical determination of the logarithmic decrement
$\delta(\theta)$ associated to the  direction $\theta$ in terms
of the singular manifold in the $(y_1,\,y_2)$-plane.}
\label{f:podaire}
\end{figure}

All this simplifies considerably for the function $F$, the solution of the
similarity equation. Indeed, all the relevant singularities have real part at
the point $(\pi,\,\pi)$ and we can restrict everything to the pure imaginary
plane $(y_1,\,y_2)$ passing through this point.  Let the
restriction\footnote{If, as is likely, the singular manifold is a complex
analytic curve, it depends on one complex parameter or on two real ones; hence
it cannot be entirely in the $(y_1,\,y_2)$-plane.} of the singular manifold to
this plane have the parametric representation $y_1(\theta),\, y_2(\theta)$,
where $y_1$ and $y_2$ are differentiable functions of the parameter $\theta$,
chosen in such a way that the angle between the $y_1$ axis and the tangent at
the singular manifold is $\theta + \pi/2$. We obviously have
\begin{equation}
    \frac{dy_2}{dy_1} =-\frac{1}{\tan \theta}.
    \label{oneovertan}
    \end{equation}
This representation is
convenient because the aformentioned  condition of having a double root 
is easily shown to express the tangency to the singular manifold of 
the straight line perpendicular to the 
direction $\theta -\pi$ at a distance $\delta(\theta)$ from the origin, as
illustrated in Fig.~\ref{f:podaire}.\footnote{The subtraction of $\pi$ is
because the wavector has negative components and thus lies in the
third quadrant $\pi <\theta <3\pi /2$.} 
\begin{figure}[ht]
\iffigs
\centerline{\psfig{file=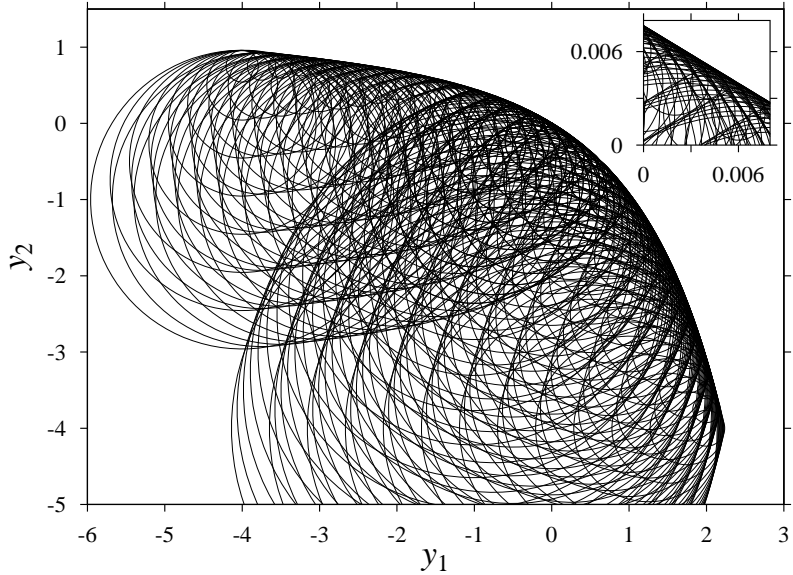,width=11.0cm,clip=}} 
\else\drawing 60 10 {analyticity disks}
\fi
\caption{The singular manifold in the $(y_1,\,y_2)$-plane constructed
as an envelope of numerically determined analyticity disks for 305 different
choices of parareal planes. Inset: enlargement near the origin.}
\label{f:envelope}
\end{figure}

The equation for the tangent reads
\begin{equation}
    \delta(\theta) =y_1 \cos (\theta -\pi) +y_2 \sin (\theta -\pi) .
    \label{deltafromy1y2}
    \end{equation}
The system \rf{oneovertan}-\rf{deltafromy1y2} is easily inverted 
(provided the singular manifold restricted to the   
$(y_1,\,y_2)$-plane is convex) to give
\begin{eqnarray}
  y_1(\theta) &=& -\delta(\theta) \cos \theta+\delta'(\theta) \sin 
  \theta   \label{singmanfromdelta1} \\
  y_2(\theta) &=& -\delta(\theta) \sin \theta -\delta'(\theta) \cos 
  \theta,
    \label{singmanfromdelta2}
\end{eqnarray}
where $\delta'(\theta) \equiv d \delta(\theta)/d\theta$.
 
 These equations allow in principle the construction of the singular
 manifold from the knowledge of the angular dependence of the
 logarithmic decrement $\delta(\theta)$, as given in Fig.~\ref{f:deltaoftheta}.
 
 In practice, because $\delta(\theta)$ changes very quickly near its minimum,
 this is not a very well conditioned procedure.  An alternative construction
 by envelope of analyticity disks is now proposed. We observe that, when we
 replace the real plane by a parareal plane shifted by an imaginary vector
 $(h_1,\,h_2)$, the location on the singular manifold nearest to this parareal
 plane is generically different from the one nearest to the original real
 plane. For a general 2D problem without special symmetry, the singular
 manifold is a one-dimensional complex manifold, and thus may be parametrized
 with two real variables, e.g. $h_1$ and $h_2$. For the special case
 considered here, we can work in the $(y_1,\,y_2)$ plane and draw around each
 point $(h_1,\,h_2)$ an analyticity disk of radius $\delta(h_1,h_2)$. The
 latter is determined as the logarithmic decrement of the shell-summed average
 of $\exp (k_1h_1+k_2h_2) \hat F_{k_1,k_2}$, where the exponential factor is
 the consequence in Fourier space of the imaginary translation
 $(ih_1,\,ih_2)$.\footnote{These imaginary shifts introduce a bias similar to
 what is done in the \citet{cramer38} derivation of the law of large
 deviations.}  It is easily shown that
\begin{equation}
\delta(h_1,h_2) = \min_\theta \, \left(\delta(\theta) -h_1\cos \theta - h_2
\sin \theta \right),
\label{pseudolegendre}
\end{equation}
where $\delta(\theta)$ is the logarithmic decrement in the direction $\theta$,
defined in Section~\ref{s:results}. Since   $\delta(\theta)$ is rather poorly
determined, it is better to measure $\delta(h_1,h_2)$ directly. We 
perform this calculation for 305 choices of the pair
 $(h_1,\,h_2)$, taken on a regular grid. The shell-summed averages are fitted
to an exponential with an algebraic prefactor. Only points displaying
decreasing exponentials, i.e., located below the singular manifold
are kept. The result is shown in Fig.~\ref{f:envelope}, where the
singular manifold emerges as the envelope of all the analyticity disks.
Let us finally observe that the method of analyticity disks can
be generalized to problems without any special symmetry; this will be 
discussed elsewhere.

\section{Concluding remarks}
\label{s:conclusion}

 Let us now summarize what we have learned about singularities for the
similarity equation \rf{asympteuler} governing the 
short-time asymptotics.  First, we remind the reader
that for one-dimensional problems, complex-space singularities can be at
isolated points but need not: for nonintegrable ODE's they often form fractal
natural boundaries \citep{jmp}. In two dimensions or higher, isolated
singularities are ruled out: the singular set is a complex manifold (or
worse). In the present case, the evidence is that the singularities are on a
smooth and probably analytic one-dimensional complex manifold.  This was
already conjectured in FMB on the basis of numerical results
for the full time-dependent Euler equation.  Now, we have an explicit 
construction of the singular manifold by the
method of envelope of analyticity disks (Section~\ref{s:singular}), which
gives strong numerical evidence for smoothness.

The very clean scaling we have observed for the Fourier transform of the
solution near the singularity (Fig.~\ref{f:ssashifted}) implies that the
vorticity is infinite on the singular manifold Since in 2D the vorticity is
conserved along Lagrangian fluid trajectories (both in the real and the
complex domain), this result strongly suggests that the singular manifold is
mapped to complex infinity by the inverse Lagrangian map.\footnote{Note that
this does not imply the absence of complex-space singularities in Lagrangian
coordinates: fluid particles situated initially at finite complex locations
can be mapped to infinity later on \citep{pm}. }

We have studied here a special flow, whose two-mode initial condition given by
\rf{deuxmodes} has a center of symmetry at $(\pi,\, \pi)$. Such symmetries are
hard to avoid when using a minimal number of Fourier modes.  At short times,
the symmetry constrains the complex-space singularity nearest to the real
domain to have the real part of its location at $(x_1, \, x_2) = (\pi,\,
\pi)$.  In FMB we have shown that this ceases to be the case at later times.
We do of course hope that there is something universal in the nature of the
singularities, be it only their very existence. We must however stress that in
the present case we have a singular manifold with a continuous piece and not
just isolated points when $(x_1, \, x_2) = (\pi,\, \pi)$. This is definitely
non-generic, since a manifold with complex dimension 1 may be viewed as a
two-dimensional real manifold in four-dimensional real space. When
higher-order harmonics are added to the initial condition \rf{deuxmodes},
the situation becomes more complicated and the short-time asymptotic equation
will in general depend on the ratio $y_2/y_1$, that is on the direction in 
which the imaginary coordinates become large. 

Let us finally point out that one particularly challenging problem is to
actually prove that there are singularities in the complex domain: at the
moment we are not aware of a single instance of a solution to the
incompressible Euler equation with smooth initial data for which the existence
of a singularity (real or complex) is demonstrated. In the present case, the
proof may be facilitated by the observation made in Section~\ref{s:results}
that, in the $(y_1,\,y_2)$-plane the vorticity is real and can be represented
as a double power series with (apparently) nonnegative coefficients. A lower
bound on these coefficient -- which are obtained from the solutions of the
recurrence relations \rf{e:beta_recursion} -- could lead to such a proof.

\vspace{4mm} \par\noindent{\bf Acknowledgments}\\[1ex] We are grateful to
Walter Pauls, Tetsuo Ueda and Vladislav Zheligovsky for useful
remarks. Computational resources were provided by the Yukawa Institute
(Kyoto). This research was supported by the European Union under contract
HPRN-CT-2000-00162 and by the Indo-French Centre for the Promotion of Advanced
Research (IFCPAR~2404-2). TM was supported by
the Grant-in-Aid for Young Scientists [(B), 15740237, 2003] and
the Grant-in-Aid for the 21st Century COE
``Center for Diversity and Universality in Physics'' from the Japanese Ministry
of Education
and received also partial support from the French Ministry of Education.
JB acknowledges support from the National Science Foundation under Agreement
No. DMS-9729992.


\begin{thebibliography}{99}
%
\bibitem[Bailey(1995)]{bailey}
Bailey,~D.H., 1995.
A fortran-90 based multiprecision system,
RNR Technical Report RNR-94-013.
See also \mbox{http://crd.lbl.gov/$\, \tilde{}\, $dhbailey/}	
%
\bibitem[Bardos, Benachour and Zerner(1976)]{bbz76}
Bardos,~C., Benachour,~S.,  Zerner,~M., 1976.
Analyticit\'{e} des solutions periodiques de l'\'{e}quation
d'Euler en deux dimensions, C. R. Acad. Sc. Paris
282 A, 995--998.
%
\bibitem[Bajer and Moffat(2003)]{zakopane}
Bajer,~K., Moffatt,~H.K. (ed), 2003. 		     
Tubes, Sheets and Singularities in Fluid Dynamics:
Proceedings of the NATO ARW, 2--7 September 2001, Zakopane, Poland, 
Kluwer Academic Publishers, Dordrecht.
%
\bibitem[Brachet \textit{et al.}(1983)]{brachetetal}
Brachet,~M.-E.,  Meiron,~D.I., Orszag,~S.A.,
Nickel,~B.G., Morf,~R.H., Frisch,~U., 1983.
Small-scale structure of the Taylor-Green vortex,
J.\ Fluid\ Mech. 130 411--452.					   
%
\bibitem[Caflisch, Hou and Lowengrub(1999)]{caflishetal}
Caflisch,~R.E., Hou,~T.Y., Lowengrub, J., 1999.
Almost optimal convergence of the point vortex method for vortex sheets
using numerical filtering, Math. Comput. 68, 1465--1496.
%
\bibitem[Carrier, Krook and Pearson(1966)]{carrier-krook-pearson} 
Carrier,~G.F., Krook,~M., Pearson,~C.E., 1966.
Functions of a complex variable: theory and technique,
McGraw-Hill, New York.		     
%
\bibitem[Chang, Tabor and Weiss(1982)]{jmp}
Chang,~Y.F., Tabor,~M., Weiss,~J., 1982.
Analytic structure of the H\'enon--Heiles hamiltonian in integrable
and nonintegrable regimes,
J. Math. Phys. 23, 531--538.
%
\bibitem[Cram\'er(1938)]{cramer38}
Cram\'er,~H., 1938.
Sur un nouveau th\'eor\`eme-limite de la th\'eorie des probabilit\'es,
Actualit\'es Scientifiques et Industrielles, 736, 5--23
%
\bibitem[Dienes(1931)]{vivanti}
Dienes,~P., 1931.
The Taylor Series, an Introduction to the Theory of Functions of a
Complex Variable, Oxford University Press. 
%
\bibitem[Frisch(1984)]{frisch84}
Frisch,~U., 1984.
The analytic structure of turbulent flows, in Proceed. Chaos and
statistical methods, Sept. 1983, Kyoto, Y.~Kuramoto, ed.
pp.~211--220, Springer.					   
%
\bibitem[Frisch, Matsumoto and Bec(2003)]{fmb03}
Frisch,~U., Matsumoto,~T., Bec,~J., 2003.
Singularities of Euler flow? Not out of the blue!
J.~Stat.~Phys. 113, 761--781.
%
\bibitem[Frisch and Morf(1981)]{frisch-morf} 
Frisch,~U.,  Morf,~R., 1981.
Intermittency in nonlinear dynamics and singularities at complex times,
Phys.~Rev.~A 23, 2673--2705.
%
\bibitem[Kida(1985)]{kida85}
Kida,~S., 1985.
Three-dimensional periodic flows with high-symmetry, 
J. Phys. Soc. Japan 54, 2132--2136.
%
\bibitem[Krasny(1986)]{krasny}
Krasny,~R., 1986.	
A study of singularity formation in a vortex sheet
by the point-vortex approximation,
J. Fluid Mech. 167, 65--93.
%
\bibitem[Majda and Bertozzi(2001)]{majda-bertozzi} 
Majda,~A.J., Bertozzi,~A.L., 2001.
Vorticity and Incompressible Flow 
(Section 9.4), 	
Cambridge Texts in Applied Mathematics,
Cambridge University Press, Cambridge.
%
\bibitem[Pauls and Matsumoto(2005)]{pm} 
Pauls,~W., Matsumoto,~T., 2005.
Lagrangian singularities of steady two-dimensional 
flow, Geophys. Astrophys. Fluid Dynam. 99, 61--75.
%
\bibitem[Pelz(1997)]{p97}
Pelz,~R.B., 1997.
Locally self-similar, finite-time collapse in a high-symmetry vortex filament model,
Phys. Rev. E 55, 1617--1626.
%
\bibitem[Pelz(2001)]{p02}	
Pelz,~R.B., 2001.
Symmetry and the hydrodynamic blow-up problem,
J. Fluid Mech. 444, 299--320.					   
%
\bibitem[Pelz and Gulak(1997)]{pg97}
Pelz,~R.B., Gulak,~Y., 1997.	
Evidence for a real-time singularity in hydrodynamics from time series analysis,
Phys. Rev. Lett. 79, 4998--5001.
%
\bibitem[Poincar\'e(1899)]{poincare}
Poincar\'{e},~H., 1899. Les M\'{e}thodes Nouvelles
de la M\'{e}canique C\'{e}leste, Gauthier-Villars, Paris
reprinted by Dover in 1957.
%
\bibitem[Sulem, Sulem and Frisch(1983)]{tracing}
Sulem,~C., Sulem,~P.-L., Frisch,~H., 1983. Tracing complex singularities with spectral methods,
J. Comput. Phys. 50, 138--161.
%
\bibitem[Tsikh(1993)]{tsikh}
Tsikh,~A., 1993.
Conditions for absolute convergence of series of Taylor coefficients of meromorphic functions of two variables, Math. USSR Sbornik 74,  336--360.
\end{thebibliography}
\end{document}